# Inductively Coupled In-Circuit Impedance Measurement and Its EMC Applications

Zhenyu Zhao, Fei Fan, Huamin Jie, Quqin Sun, Pengfei Tu, Wensong Wang, and Kye Yak See

*Abstract*—In-circuit impedance provides key information for many EMC applications. The inductive coupling approach is a promising method for in-circuit impedance measurement because its measurement setups have no direct electrical contact with the energized system under test, thus greatly simplifying the on-site implementation. This paper presents and summaries the latest research on the inductive coupling approach and its EMC applications. First of all, three common measurement setups for this approach and their respective pros and cons are discussed. Subsequently, their EMC applications are introduced. Finally, recommendations for future research are listed.

## I. INTRODUCTION

In-circuit impedance is a critical parameter for many EMC applications [1]-[6]. In general, there are three approaches for in-circuit impedance measurement, namely the voltage-current (V-I) approach, capacitive coupling approach, and inductive coupling approach [7]. The V-I approach measures the voltage across and current flow of the system under test (SUT) and incorporates signal processing algorithms to extract in-circuit impedance [8]-[10]. The capacitive coupling approach employs coupling capacitors to block low-frequency power supply but to pass high-frequency test signals so that the impedance analyzer (IA) or vector network analyzer (VNA) connected to coupling capacitors can perform in-circuit impedance measurement [11]-[13]. The measurement setups for both approaches require some form of electrical contact with the energized SUT and hence may present electrical hazards. In contrast, the inductive coupling approach eliminates direct electrical contact with the energized SUT, thereby greatly simplifying on-site implementation [14].

To measure the in-circuit impedance of one SUT via the inductive coupling approach, three measurement setups are commonly used, namely the frequency-domain two-probe setup (FD-TPS) [15]-[18], time-domain two-probe setup (TD-TPS) [19]-[22], and frequency-domain single-probe setup (FD-SPS) [23]-[25]. Although a frequency-domain three-probe setup [26] and a time-domain multi-probe setup [27] were also designed, they are used to simultaneously measure the in-circuit impedances of SUTs on two and multiple branches powered by the same power source, respectively.

This research work was supported by Nanyang Technological University. (Corresponding author: Fei Fan)

The authors are with the School of Electrical and Electronics Engineering, Nanyang Technological University, Singapore 639798. (email: zhao0245@e.ntu.edu.sg, fanf0003@e.ntu.edu.sg, ekysee@ntu.edu.sg).

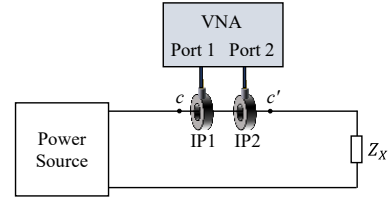

Fig. 1. In-circuit impedance measurement through FD-TPS.

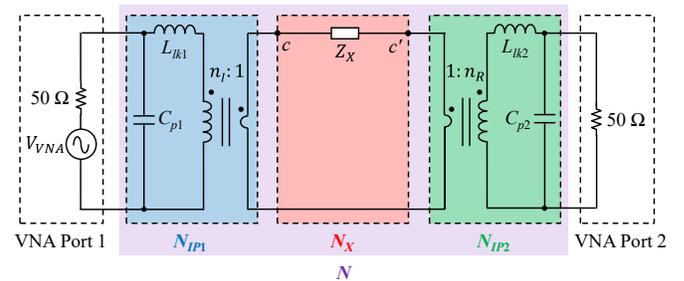

Fig. 2. Cascaded two-port network equivalent circuit of Fig. 1.

In this paper, the latest research on the inductive coupling approach as well as its EMC applications are presented and summarized. Firstly, three common measurement setups (i.e., FD-TPS, TD-TPS, and FD-SPS) for this approach and their respective pros and cons are discussed. Then, their EMC applications are introduced and the following two applications are elaborated in particular, namely the radiated emission estimation of photovoltaic (PV) systems and the noise source impedance extraction at the AC input of variable frequency drive (VFD) systems. Finally, recommendations for future research on this approach are listed.

## II. MEASUREMENT SETUPS FOR INDUCTIVE COUPLING APPROACH

### A. Frequency-Domain Two-Probe Setup

As shown in Fig. 1, the FD-TPS usually includes two clamp-on inductive probes (IP1 and IP2) and a VNA. For measuring the in-circuit impedance of the SUT ($Z_X$) that is energized by a power source, the IP1 and IP2 are clamped on the wiring connection of the SUT with the clamping position represented as *c-c'*. One port of the VNA injects a stepped swept-sine excitation signal into the SUT via an inductive probe and the other port of the VNA measures the response of the signal via another inductive probe. Fig. 2 shows its cascaded two-port network equivalent circuit. $N_{IPi}$ ($i$ = 1, 2) represents the two-port network of the IP$i$ with the wiring being clamped, where $L_{lki}$ and $C_{pi}$ represent the leakage inductance and parasitic capacitance between the winding of the IP$i$ and its frame, respectively [28]. $N_X$ is the two-port network of the SUT. Since $N_{IP1}$, $N_X$, and $N_{IP2}$ are cascaded, the resultant two-port network $N$ can be expressed as:

$$N = N_{IP1} N_X N_{IP2} \quad (1)$$

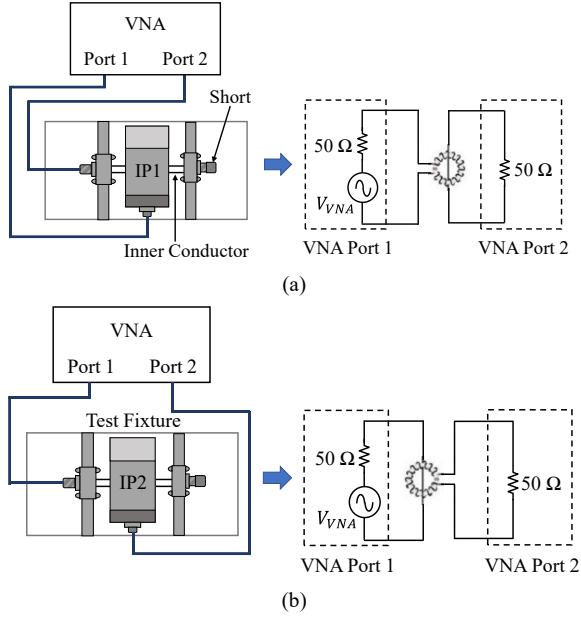

Fig. 3. Test fixture to characterize (a) $N_{IP1}$ and (b) $N_{IP2}$.

Expressing these two-port networks in terms of transmission (ABCD) parameters, (1) can be rewritten as:

$$\begin{bmatrix} A & B \\ C & D \end{bmatrix} = \begin{bmatrix} A_{IP1} & B_{IP1} \\ C_{IP1} & D_{IP1} \end{bmatrix} \begin{bmatrix} A_X & B_X \\ C_X & D_X \end{bmatrix} \begin{bmatrix} A_{IP2} & B_{IP2} \\ C_{IP2} & D_{IP2} \end{bmatrix} \quad (2)$$

By solving $N_X$, $Z_X$ can be obtained since $Z_X = B_X$ [29]. From (2), the ABCD parameters of $N_X$ can be derived when $N$, $N_{IP1}$, and $N_{IP2}$ are known. Among them, the ABCD parameters of $N$ can be derived from the measured scattering (S) parameters using the VNA. The correlation of the ABCD parameters and S-parameters is given by [29]:

$$\begin{aligned} A &= \frac{(1+S_{11})(1-S_{22}) + S_{12}S_{21}}{2S_{21}} \\ B &= Z_0 \frac{(1+S_{11})(1+S_{22}) - S_{12}S_{21}}{2S_{21}} \\ C &= \frac{1}{Z_0} \frac{(1-S_{11})(1-S_{22}) - S_{12}S_{21}}{2S_{21}} \\ D &= \frac{(1-S_{11})(1+S_{22}) + S_{12}S_{21}}{2S_{21}} \end{aligned} \quad (3)$$

where $Z_0$ is the reference impedance, and $Z_0 = 50$ Ω.

To extract the ABCD parameters of $N_{IP1}$ and $N_{IP2}$, a specific test fixture shown in Fig. 3 is employed. The inductive probe (IP1 or IP2) is clamped on the inner conductor of the test fixture and the outer conductor of the test fixture acts as the common reference return path for the VNA. One end of the test fixture is terminated with a short to make the inner conductor shorted with the outer conductor. To obtain $N_{IP1}$, the VNA Port 1 connects to the IP1 and its Port 2 connects to the other end of the test fixture, as shown in Fig. 3(a). To obtain $N_{IP2}$, the VNA Port 1 connects to the other end of the test fixture and its Port 2 connects to the IP2, as shown in Fig. 3(b). In this way, the S-parameters of $N_{IP1}$ and $N_{IP2}$ can be measured directly using the VNA, and finally, their ACBD parameters can be derived from the conversion of the measured S-parameters. After $N_{IP1}$ and $N_{IP2}$ are determined, $Z_X$ can be obtained via the direct measurement of $N$ using the VNA.

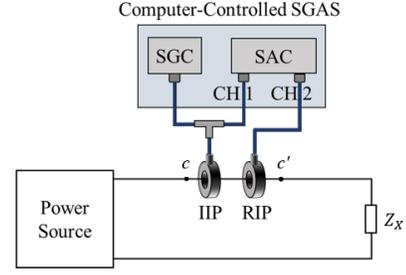

Fig. 4. In-circuit impedance measurement through TD-TPS.

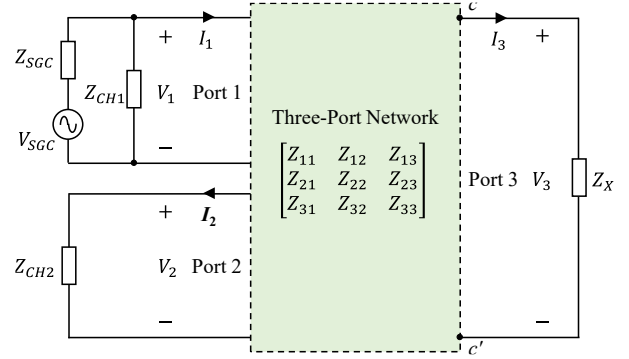

Fig. 5. Three-port network equivalent circuit of Fig. 4.

B. Time-Domain Two-Probe Setup

The FD-TPS is often used to measure the time-invariant in-circuit impedance. To measure the time-variant in-circuit impedance, the TD-TPS was developed. As shown in Fig. 4, the TD-TPS usually consists of one clamp-on injecting inductive probe (IIP), one clamp-on receiving inductive probe (RIP), and a computer-controlled signal generation and acquisition system (SGAS). The signal generation card (SGC) of the SGAS generates a single-sine excitation signal and the signal is injected into the SUT via the IIP. The RIP is used to monitor the response of the signal. Channel 1 (CH1) and Channel 2 (CH2) of the signal acquisition card (SAC) of the SGAS measure the excitation signal voltage at the IIP and the response signal voltage at the RIP, respectively. Fig. 5 shows the three-port network equivalent circuit of Fig. 4. Compared with the cascaded two-port network equivalent circuit, the three-port equivalent circuit takes into account the probe-to-probe coupling effect [20].

In Fig. 5, $V_{SGC}$ and $Z_{SGC}$ represent the equivalent source voltage and impedance of the SGC, respectively. $Z_{CH1}$ and $Z_{CH2}$ represent the internal impedances of CH1 and CH2 of the SAC, respectively. $V_1$ and $I_1$ denote the excitation signal voltage and current at the IIP, respectively. $V_2$ and $I_2$ denote the response signal voltage and current at the RIP, respectively; where $I_2 = V_2/Z_{CH2}$. Note that $V_1$ and $V_2$ can be directly extracted by the SAC. $V_3$ represents the induced signal voltage between $c$ and $c'$, and $I_3$ represents the induced signal current passing through $Z_X$ to be measured, where $I_3 = V_3/Z_X$. $Z_{ij}$ ($i, j = 1, 2, 3$) are the impedance parameters of the three-port network. Based on the three-port network, the above-mentioned voltages and currents are related by:

$$\begin{bmatrix} V_1 \\ V_2 \\ V_3 \end{bmatrix} = \begin{bmatrix} Z_{11} & Z_{12} & Z_{13} \\ Z_{21} & Z_{22} & Z_{23} \\ Z_{31} & Z_{32} & Z_{33} \end{bmatrix} \begin{bmatrix} I_1 \\ I_2 \\ I_3 \end{bmatrix} \quad (4)$$

Substituting $I_2 = V_2/Z_{CH2}$ and $I_3 = V_3/Z_X$ into (4), and dividing $V_2$ at both sides, we obtain:

$$\begin{bmatrix} V_1/V_2 \\ 1 \\ V_3/V_2 \end{bmatrix} = \begin{bmatrix} Z_{11} & Z_{12} & Z_{13} \\ Z_{21} & Z_{22} & Z_{23} \\ Z_{31} & Z_{32} & Z_{33} \end{bmatrix} \begin{bmatrix} I_1/V_2 \\ 1/Z_{CH2} \\ (V_3/V_2)/Z_X \end{bmatrix} \quad (5)$$

By solving (5), $V_1/V_2$ is expressed in terms of $Z_{ij}$, $Z_{CH2}$, and $Z_X$ as follows:

$$\frac{V_1}{V_2} = \frac{a_1 \cdot Z_X + a_2}{Z_X + a_3} \quad (6)$$

where

$$a_1 = \frac{Z_{11}}{Z_{21}} \cdot \left(1 - \frac{Z_{22}}{Z_{CH2}}\right) + \frac{Z_{12}}{Z_{CH2}}$$

$$a_2 = \left(Z_{13} - \frac{Z_{11}Z_{23}}{Z_{21}}\right) \cdot \left[\frac{Z_{31}}{Z_{21}} \cdot \left(1 - \frac{Z_{22}}{Z_{CH2}}\right) + \frac{Z_{32}}{Z_{CH2}}\right] \quad (7)$$
$$- \left(Z_{33} - \frac{Z_{31}Z_{23}}{Z_{21}}\right) \cdot \left[\frac{Z_{11}}{Z_{21}} \cdot \left(1 - \frac{Z_{22}}{Z_{CH2}}\right) + \frac{Z_{12}}{Z_{CH2}}\right]$$

$$a_3 = \frac{Z_{31}Z_{23}}{Z_{21}} - Z_{33}$$

Finally, $Z_X$ is denoted as a function of $V_1/V_2$ as follows:

$$Z_X = \frac{a_3 \cdot (V_1/V_2) - a_2}{-V_1/V_2 + a_1} \quad (8)$$

From (8), $Z_X$ can be obtained via the measurement of $V_1$ and $V_2$ once $a_1$, $a_2$ and $a_3$ are known. Considering $a_1$, $a_2$ and $a_3$ are determined by $Z_{ij}$ and $Z_{CH2}$ that keep unchanged for a given TD-TPS, a calibration technique [21] is applied to determine their values. To execute this calibration, three distinct known calibration components (CC1, CC2 and CC3) with their respective impedances ($Z_{CC1}$, $Z_{CC2}$ and $Z_{CC3}$, where $Z_{CC1} \neq Z_{CC2} \neq Z_{CC3}$) are needed. The values of $Z_{CC1}$, $Z_{CC2}$ and $Z_{CC3}$ are extracted with a precision IA. Once $Z_{CC1}$, $Z_{CC2}$ and $Z_{CC3}$ are obtained and based on (8), when $c$-$c'$ is respectively terminated with $Z_{CC1}$, $Z_{CC2}$ and $Z_{CC3}$, we obtain

$$Z_{CC1} = \frac{a_3(V_1/V_2|_{CC1}) - a_2}{-V_1/V_2|_{CC1} + a_1} \quad (9)$$

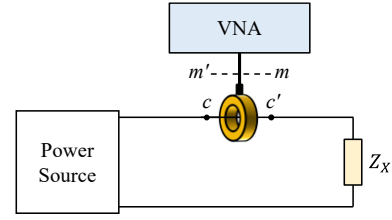

Fig. 6. In-circuit impedance measurement through FD-SPS.

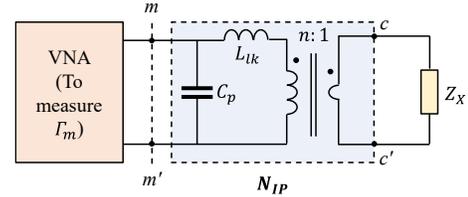

Fig. 7. Two-port network equivalent circuit of Fig. 6.

$$Z_{CC2} = \frac{a_3(V_1/V_2|_{CC2}) - a_2}{-V_1/V_2|_{CC2} + a_1} \quad (10)$$

$$Z_{CC3} = \frac{a_3(V_1/V_2|_{CC3}) - a_2}{-V_1/V_2|_{CC3} + a_1} \quad (11)$$

where $V_1/V_2|_{CC_i}$ ($i = 1, 2, 3$) can be measured directly using the TD-TPS. Based on (9)-(11), $a_1$, $a_2$ and $a_3$ are finally solved as:

$$a_1 = [(V_1/V_2|_{CC1})(V_1/V_2|_{CC2})(Z_{CC1} - Z_{CC2})$$
$$+ (V_1/V_2|_{CC1})(V_1/V_2|_{CC3})(Z_{CC3} - Z_{CC1})$$
$$+ (V_1/V_2|_{CC2})(V_1/V_2|_{CC3})(Z_{CC2} - Z_{CC3})]/\Lambda$$

$$a_2 = [(V_1/V_2|_{CC1})(V_1/V_2|_{CC2})(Z_{CC2} - Z_{CC1})Z_{CC3}$$
$$+ (V_1/V_2|_{CC1})(V_1/V_2|_{CC3})(Z_{CC1} - Z_{CC3})Z_{CC2}$$
$$+ (V_1/V_2|_{CC2})(V_1/V_2|_{CC3})(Z_{CC3} - Z_{CC2})Z_{CC1}]/\Lambda \quad (12)$$

$$a_3 = [(V_1/V_2|_{CC1})(Z_{CC2} - Z_{CC3})Z_{CC1}$$
$$+ (V_1/V_2|_{CC2})(Z_{CC3} - Z_{CC1})Z_{CC2}$$
$$+ (V_1/V_2|_{CC3})(Z_{CC1} - Z_{CC2})Z_{CC3}]/\Lambda$$

$$\Lambda = (V_1/V_2|_{CC1})(Z_{CC3} - Z_{CC2}) + (V_1/V_2|_{CC2})$$
$$\cdot (Z_{CC1} - Z_{CC3}) + (V_1/V_2|_{CC3})(Z_{CC2} - Z_{CC1})$$

where $a_1$ is unit-less, $a_2$ and $a_3$ have units of ohm ($\Omega$), and $\Lambda$ has to be nonzero for $a_1$, $a_2$ and $a_3$ to be unique.

Once $a_1$, $a_2$ and $a_3$ are determined through the calibration, $Z_X$ can be derived from (8) through the measured $V_1$ and $V_2$. To measure $V_1$ and $V_2$ continuously, their time-domain counterparts $v_1$ and $v_2$ are sampled and a moving window discrete Fourier transform algorithm is employed. Based on the continuously measured $V_1$ and $V_2$, the time-variant $Z_X$ can be extracted [19].

### C. Frequency-Domain Single-Probe Setup

The FD-SPS does not need to consider the probe-to-probe coupling issue encountered by the FD-TPS and TD-TPS as only single probe is used in the FD-SPS. As shown in Fig. 6, it usually consists of one clamp-on inductive probe and a frequency-domain measurement instrument (VNA). To measure $Z_X$, the VNA generates a stepped swept-sine excitation through its signal source and provides the reflection coefficient at $m$-$m'$ represented as $\Gamma_m$. Fig. 7 shows the two-port network equivalent circuit of Fig. 6, where $N_{IP}$ denotes the two-port network of the inductive probe with the wire being clamped. $L_{lk}$ and $C_p$ represent the leakage inductance and parasitic capacitance between the winding of the probe and its frame, respectively [28]. Note that the parasitic parameters of the inductive probe have been included in $N_{IP}$. Using the *ABCD* parameters to represent $N_{IP}$, the relationship between $Z_X$ and $\Gamma_m$ is established by:

$$Z_X = \frac{k_1 \cdot \Gamma_m + k_2}{\Gamma_m + k_3} \quad (13)$$

where

$$k_1 = -\frac{Z_0 \cdot D + B}{Z_0 \cdot C + A}$$
$$k_2 = -\frac{Z_0 \cdot D - B}{Z_0 \cdot C + A} \quad (14)$$
$$k_3 = \frac{Z_0 \cdot C - A}{Z_0 \cdot C + A}$$

where $Z_0$ denotes the reference impedance of the VNA.

From (13), $Z_X$ can be obtained via the measurement of $\Gamma_m$ once $k_1$, $k_2$ and $k_3$ are known. Considering $k_1$, $k_2$ and $k_3$ are determined by the *ABCD* parameters of $N_{IP}$ and $Z_0$, which remain unchanged for a given FD-SPS, the calibration introduced described in Subsection B also can be used to determine the values of $k_1$, $k_2$ and $k_3$. The calibration components and their respective impedances are still denoted as CC$i$ and $Z_{CCi}$ ($i$ = 1, 2, 3), respectively. After executing this calibration, $k_1$, $k_2$ and $k_3$ can be finally solved as:

$$\begin{aligned}
k_1 &= [Z_1(Z_2 - Z_3)\Gamma_m|_{Z_1} \\
&\quad + Z_2(Z_3 - Z_1)\Gamma_m|_{Z_2} \\
&\quad + Z_3(Z_1 - Z_2)\Gamma_m|_{Z_3}]/\Delta \\
k_2 &= [Z_3(Z_1 - Z_2)\Gamma_m|_{Z_1}\Gamma_m|_{Z_2} \\
&\quad + Z_2(Z_3 - Z_1)\Gamma_m|_{Z_1}\Gamma_m|_{Z_3} \\
&\quad + Z_1(Z_2 - Z_3)\Gamma_m|_{Z_2}\Gamma_m|_{Z_3}]/\Delta \\
k_3 &= [(Z_1 - Z_2)\Gamma_m|_{Z_1}\Gamma_m|_{Z_2} \\
&\quad + (Z_3 - Z_1)\Gamma_m|_{Z_1}\Gamma_m|_{Z_3} \\
&\quad + (Z_2 - Z_3)\Gamma_m|_{Z_2}\Gamma_m|_{Z_3}]/\Delta \\
\Delta &= \Gamma_m|_{Z_1}(Z_2 - Z_3) \\
&\quad + \Gamma_m|_{Z_2}(Z_3 - Z_1) \\
&\quad + \Gamma_m|_{Z_3}(Z_1 - Z_2)
\end{aligned} \quad (15)$$

where $\Gamma_m|_{Z_i}$ ($i$ = 1, 2, 3) represents the measured reflection coefficient using the VNA at $m$-$m'$ for the respective $Z_i$. $k_1$ and $k_2$ have units of ohm (Ω), $k_3$ is unit-less, and Δ has to be nonzero for $k_1$, $k_2$ and $k_3$ to be unique.

### D. Comparison of Measurement Setups

Table I compares the above three measurement setups (i.e., FD-TPS, TD-TPS, and FD-SPS) in four aspects: excitation signal type, type of measured parameters, capability of time-variant measurements, and whether suffers from the probe-to-probe coupling. From the table, both FD-TPS and FD-SPS perform measurements through stepped swept-sine excitation while TD-TPS through single-sine excitation. Besides, the FD-TPS and FD-SPS extract the in-circuit impedance by measuring *S*-parameters while the TD-TPS extracts the in-circuit impedance by measuring time-domain voltages. Therefore, the FD-TPS and FD-SPS are often used for time-invariant in-circuit impedance measurement. In contrast, the TD-TPS not only can be used for time-invariant but also time-variant in-circuit impedance measurement. However, since the TD-TPS performs the measurement through single-sine excitation, it is rather laborious to measure the in-circuit impedance in a wide frequency range compared to the FD-TPS and FD-SPS. In addition to the above, compared to the FD-TPS and TD-TPS, the FD-SPS eliminates the need to account for the probe-to-probe coupling effect as only single inductive probe is used in this measurement setup.

TABLE I
COMPARISON OF MEASUREMENT SETUPS OF INDUCTIVE COUPLING APPROACH

| Aspects | FD-TPS | TD-TPS | FD-SPS |
|---|---|---|---|
| Excitation signal | Stepped swept-sine | Single-sine | Stepped swept-sine |
| Measured parameters | *S*-parameters | Time-domain voltages | *S*-parameters |
| Time-variant measurements | Inappropriate | Appropriate | Inappropriate |
| Probe-to-probe coupling | Have | Have | Don't Have |

It should be noted that for the applications where strong electrical noise and power surges are present, all the three measurement setups can incorporate a signal amplification and protection (SAP) module to improve its signal-to-noise ratio (SNR) and enhance its ruggedness. The SAP modules of the FD-TPS, TD-TPS and FD-SPS have been elaborated in references [18], [22] and [23], respectively, and will not be repeated here.

### III. EMC APPLICATIONS

The inductive coupling approach has been used in many EMC applications. For instance, the FD-TPS was originally used to extract the in-circuit impedance of power lines [15], it was later refined and applied to the design of electromagnetic interference (EMI) filters for power converters [16], and it was recently used to estimate the radiated emissions of PV systems [30]. In addition, the TD-TPS was first reported for time-variant in-circuit impedance monitoring of switching circuits [19], later for voltage-dependent capacitance extraction of power semiconductor devices [3], and more recently for online detection of stator insulation faults in inverter-fed induction motors through real-time CM impedance monitoring [22]. It is worth mentioning that the TD-TPS is very potential in condition monitoring applications due to its time-variant measurement capability. For the FD-SPS, it was recently developed and used for noise source impedance extraction at the AC input of VFD systems [24], [25]. Among the various EMC applications of the inductive coupling approach mentioned above, the following two applications will be elaborated in particular, namely the radiated emission estimation of PV systems and the noise source impedance extraction at the AC input of VFD systems.

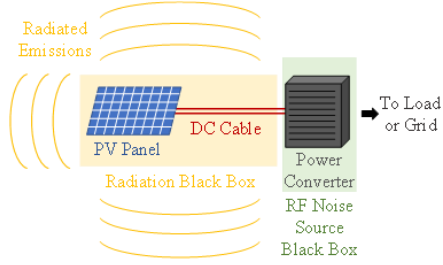

Fig. 8. Schematic diagram of a PV system.

TABLE II
Specifications of Components in the PV System

| Components | Specifications |
| --- | --- |
| PV Panel | Nominal Peak Power: 60 W<br>Nominal Voltage: 18.3 V<br>Nominal Current: 3.27 A<br>Dimensions: 685 mm × 670 mm ×35 mm |
| Power Converter | DC-DC Buck Converter<br>Input: 10-40 V, 5 A |
| DC Cable | Conductor Material: Copper<br>Conductor Radius: 1.2 mm |

### A. Radiated Emission Estimation of Photovoltaic Systems

Fig. 8 shows a basic PV system with a PV panel that converts light energy into electrical energy, which is then fed to a power converter through connected DC cable to provide a regulated and stable power source. In the PV system, the power converter is a well-known source of EMI, which can radiate electromagnetic fields through the PV panel with connected DC cable [31]. Several numerical simulation-based methods have been reported to estimate the radiated emissions in different frequency ranges: 150 kHz – 30 MHz [31]-[33], 30 MHz – 300 MHz [34], [35], and 30 MHz – 1 GHz [36]. However, the power converter is not included in their models. Besides, the construction of these models requires the design details of the PV panel, which are usually not available to the users due to intellectual property protection. To overcome these limitations, a black box method employing the FD-TPS was developed to estimate the radiated emissions from a PV system in the frequency range of 150 kHz to 30 MHz [30].

For this method, the power converter is considered a "radio frequency (RF) noise source black box", represented by an equivalent noise source with an internal source impedance. In this way, the noise current output from the power converter can be easily estimated. In addition, the PV panel with connected DC cable are considered as a "radiation black box", represented by a transfer function related to the radiated emissions and the noise current output from the power converter. By combing the first and second black boxes, the radiated emissions of the PV system can be estimated. For illustration purposes, a commercially available PV system was selected as a case study. Table II lists the detailed specifications of three key components in the PV system. Since the radiated emissions of a PV system are mainly caused by the CM noise generated by the power converter [37], the RF noise source black box focuses on extracting the equivalent CM noise source of the power converter.

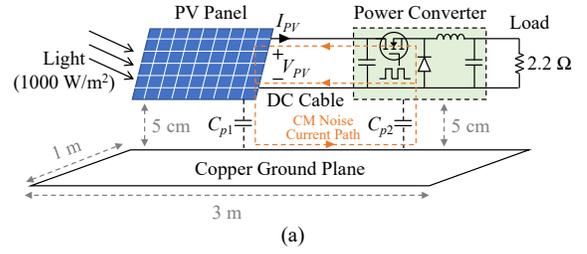

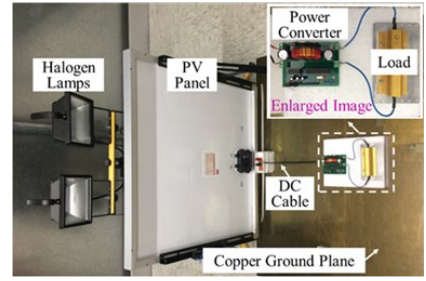

Fig. 9. Experimental Setup. (a) Schematic diagram. (b) Actual photo.

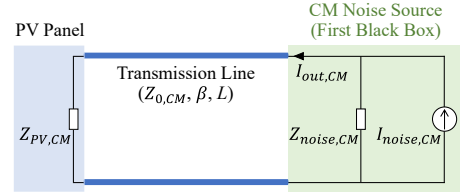

Fig. 10. CM noise equivalent circuit of the PV system.

Fig. 9 shows the experimental setup. The PV panel is illuminated by halogen lamps with a light intensity of 1000 W/m$^2$. At the given light intensity, the output DC voltage ($V_{PV}$) and current ($I_{PV}$) of the PV panel were measured to be 18.1 V and 2.6 A, respectively. The switching frequency and duty cycle of the power converter are 50 kHz and 0.55, respectively. $C_{p1}$ and $C_{p2}$ represent the parasitic capacitance between the PV panel and ground plane and the parasitic capacitance between the power converter and ground plane, respectively. The CM noise current flows in both positive and negative lines of the DC cable to the PV panel, and returns through the parasitic capacitances and the ground plane, radiating electromagnetic emissions [38].

Fig. 10 shows the CM noise equivalent circuit of the PV system, where the power converter is represented by a noise current source ($I_{noise,CM}$) with source impedance ($Z_{noise,CM}$). The DC cable connected between the power converter and the PV panel with reference to the ground plane is modelled as a transmission line, in which $Z_{0,CM}$, $\beta$ and $L$ represent its CM characteristic impedance, phase constant and length, respectively. $Z_{PV,CM}$ denotes the CM impedance of the PV panel. $I_{out,CM}$ denotes the CM noise current output from the power converter.

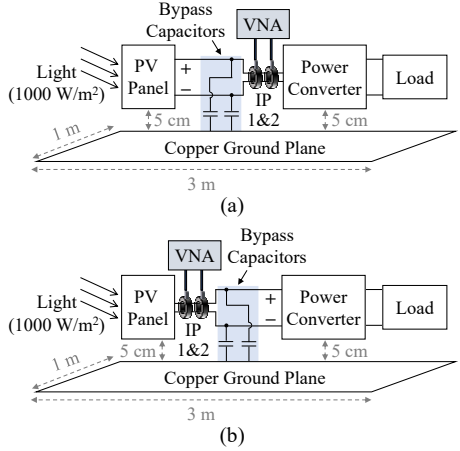

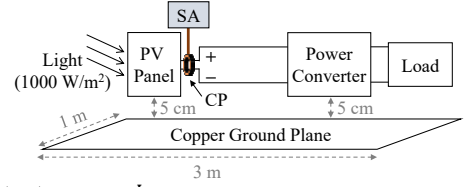

Fig. 13. Setup to measure $I_{PV,CM}$.

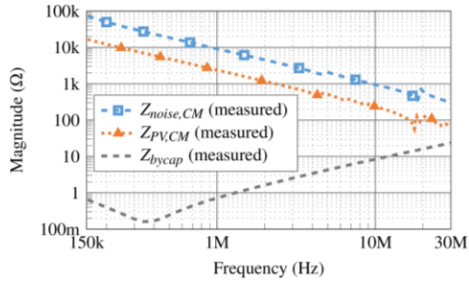

Fig. 11. In-circuit extraction of (a) $Z_{noise,CM}$ and (b) $Z_{PV,CM}$ through FD-TPS.

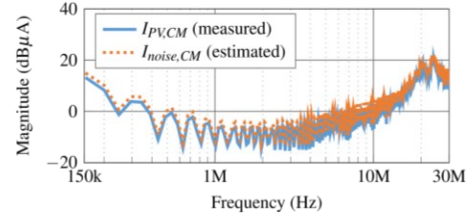

Fig. 14. Measured $I_{PV,CM}$ and estimated $I_{noise,CM}$ from 150 kHz – 30 MHz.

To model the first black box (i.e., RF noise source black box), the in-circuit $Z_{noise,CM}$ and $Z_{PV,CM}$ have to be extracted. A FD-TPS with a Bode 100 VNA and two Solar 9144-1N clamp-on inductive probes is chosen for this extraction. Fig. 11 shows the details of extracting the in-circuit $Z_{noise,CM}$ and $Z_{PV,CM}$. To extract $Z_{noise,CM}$, the two inductive probes (IP1 and IP2) are clamped on both positive and negative lines of the DC cable at the power converter side. The cable length is kept as short as possible (25 cm) to give just sufficient space for clamping IP1 and IP2, thereby minimizing the effect of cable impedance on the measurement. Additionally, two bypass capacitors are connected between each line of the DC cable and the ground, providing a low impedance path for RF test signals over the frequency range of 150 kHz – 30 MHz. Therefore, the PV panel is actually bypassed to extract $Z_{noise,CM}$. Similarly, to extract $Z_{PV,CM}$, IP1 and IP2 are clamped on both positive and negative lines of the DC cable at the PV panel side, and bypass capacitors are used to bypass the power converter for RF test signals. In this case study, the values of the bypass capacitors were chosen to be 1 μF. Fig. 12 shows the extracted $Z_{noise,CM}$ and $Z_{PV,CM}$. In addition, the impedance of the bypass capacitors ($Z_{bycap}$) was measured offline with an Agilent 4294A impedance analyzer. It is observed from Fig. 12 that $Z_{bycap}$ is significantly smaller than $Z_{noise,CM}$ and $Z_{PV,CM}$, which confirms its effectiveness as a bypass. Moreover, both $Z_{noise,CM}$ and $Z_{PV,CM}$ are capacitive in nature and can be modelled as 20-pF and 70-pF capacitors, respectively.

Based on the transmission line theory, if $I_{PV,CM}$ is known, $I_{noise,CM}$ can be estimated by [29]:

$$I_{noise,CM} = I_{PV,CM} \left[ \left( \frac{Z_{PV,CM}}{Z_{noise,CM}} + 1 \right) \cos(\beta L) + j \left( \frac{Z_{0,CM}}{Z_{noise,CM}} + \frac{Z_{PV,CM}}{Z_{0,CM}} \right) \sin(\beta L) \right] \quad (16)$$

In (16), $Z_{noise,CM}$ and $Z_{PV,CM}$ can be directly extracted using the FD-TPS. $Z_{0,CM}$, $\beta$ and $L$ are known for a given cable configuration. For a two-conductor cable above a ground plane, $Z_{0,CM}$ can be obtained from [39]:

$$Z_{0,CM} = 30\ln\left[\frac{2h}{r}\sqrt{1+\left(\frac{2h}{d}\right)^2}\right] \quad (17)$$

where $h$ represents the distance between the conductor and the ground plane, $r$ denotes the conductor radius, and $d$ represents the center-to-center distance between the two conductors. In this case study, $h$ = 50 mm, $r$ = 1.2 mm, and $d$ = 2.6 mm. Therefore, $Z_{0,CM}$ is calculated to be 242 Ω. Besides, $\beta$ can be derived from [29]:

$$\beta = 2\pi f \sqrt{\mu \varepsilon} \quad (18)$$

where $f$ represents the frequency, $\mu$ and $\varepsilon$ denote the permeability and permittivity of the medium, respectively.

Fig. 13 shows the setup to measure $I_{PV,CM}$ using a Solar 9134-1 current probe (CP) and a Rohde & Schwarz FSH4 spectrum analyzer (SA). For a DC cable length of 25 cm, Fig. 14 shows the measured $I_{PV,CM}$ and estimated $I_{noise,CM}$. With $I_{noise,CM}$ and $Z_{noise,CM}$ known, the first black box is established. Thus, $I_{out,CM}$ in the case of any cable length ($L$) can be estimated by [29]:

$$I_{out,CM} = \frac{I_{noise,CM}}{\frac{Z_{0,CM}}{Z_{noise,CM}} \cdot \frac{Z_{PV,CM} + jZ_{0,CM}\tan(\beta L)}{Z_{0,CM} + jZ_{PV,CM}\tan(\beta L)} + 1} \quad (19)$$

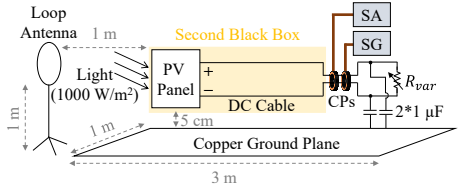

Fig. 15. Setup to extract $K$.

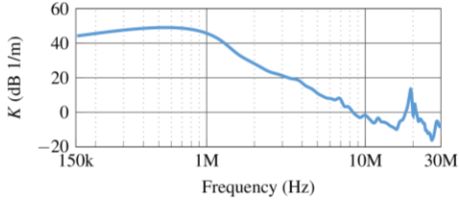

Fig. 16. Extracted $K$ from 150 kHz – 30 MHz for the PV panel with connected DC cable (3 m).

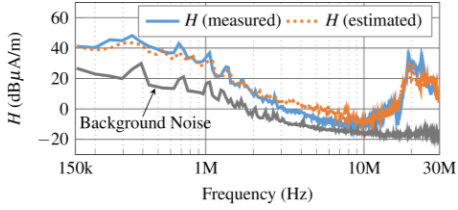

Fig. 17. Estimated and measured radiated emissions from the PV system (with 3 m DC cable) from 150 kHz – 30 MHz.

To model the second black box (i.e., radiation black box), the transfer function between the radiated emissions and $I_{out,CM}$ is established:

$$K[\text{dB 1/m}] = H[\text{dB}\mu\text{A/m}] - I_{out,CM}[\text{dB}\mu\text{A}] \quad (20)$$

where $H$ represents the radiated magnetic field in dBμA/m, and $I_{out,CM}$ is in dBμA.

Fig. 15 shows the setup to extract $K$. The DC cable length is selected to be 3 m. A variable resistor $R_{var}$ connects to the PV panel through the DC cable to make the PV panel operate at a specific operating point (i.e., $V_{PV}$ = 18.1 V and $I_{PV}$ = 2.6 A). A CP is clamped on the DC cable bundle at the variable resistor side, and then a signal generator (SG) is connected to the CP to inject CM current into the PV panel and DC cable. The injected CM current is to emulate $I_{out,CM}$. Two 1-μF capacitors are connected between each line of the DC cable and the ground plane at the variable resistor side to provide a closed CM current path for the injected RF test signals. Another CP is clamped on the same DC cable bundle at the variable resistor side, and then a SA is connected to the CP to measure $I_{out,CM}$. To measure $H$, a loop antenna (EMCO 6509), a preamplifier, and a SA were used [40]. The loop antenna is placed at 1 m height and located 1 m away from the PV panel. The whole setup is placed in a semi-anechoic chamber, and the emitted magnetic field at each emission frequency is obtained by [41]:

$$H[\text{dB}\mu\text{A/m}] = V_{rcv}[\text{dB}\mu\text{V}] - G[\text{dB}] + AF[\text{dB 1/}\Omega\text{m}] \quad (21)$$

where $V_{rcv}$ represents the signal measured using the SA in dBμV, $G$ denotes the preamplifier gain in dB, and $AF$ represents the calibrated antenna factor of the loop antenna in dB 1/Ωm. Once $H$ and $I_{out,CM}$ are obtained, $K$ can be derived from (20). Fig. 16 shows the extracted $K$ for the PV panel with connected DC cable (3 m).

By combing the estimated $I_{out,CM}$ from the RF noise source black box and the determined $K$ from the radiation black box, the radiated emissions from the PV system can be estimated. Fig. 17 shows the estimated and measured radiated emissions from the PV system (with 3 m DC cable), as well as the background noise. As shown, the background noise is much lower than the radiated emissions. In addition, the estimated radiated emissions show good agreement with the measured results over the entire frequency range of 150 kHz – 30 MHz. Hence, the effectiveness of the black box method is verified.

### B. Noise Source Impedance Extraction at the AC Input of Variable Frequency Drive Systems

The switching of power semiconductor devices in a VFD system generates conducted EMI noise that can propagate from its AC input to the power grid, affecting the normal operation of other grid-connected electrical assets [42]. Conducted EMI noise is usually divided into CM and differential-mode (DM) components. To evaluate these components, the respective CM and DM noise models of the VFD system need to be constructed [43]. Since these noise models are usually represented by respective CM and DM equivalent noise sources with internal impedances, it is necessary to extract the noise source impedances.

This subsection introduces the application of the FD-SPS to extract the noise source impedances at the AC input of a VFD system. As shown in Fig. 18, a typical VFD system includes a variable frequency controller and an induction motor with cables in between. The AC input of the VFD system is usually 3-phase or single-phase depending on the application scale. Fig. 19 shows the block diagram of measuring the noise source impedances at the AC input of the VFD system through the FD-SPS, where the VFD system is connected to the AC power through a line impedance stabilization network (LISN), where the LISN is to provide a stable and well-defined impedance at the AC power side and to prevent test signals from leaking into the power grid, affecting the operation of grid-connected sensitive electrical devices. Since the VFD system usually suffers from significant background noise and experiences power surges, the FD-SPS incorporates a SAP module to improve its SNR and enhance its ruggedness. As shown in Fig. 19(a), for in-circuit extraction of the CM impedance ($Z_{CM,VFD}$), the inductive probe is clamped on the ground cable. As shown in Fig. 19(b), for in-circuit extraction of the DM impedance ($Z_{DM,VFD}$), the ground connection between the VFD system and the LISN is left open, and the inductive probe is clamped on one of the power cables.

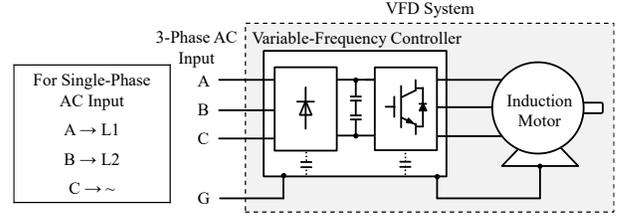

Fig. 18. Schematic diagram of a typical VFD system.

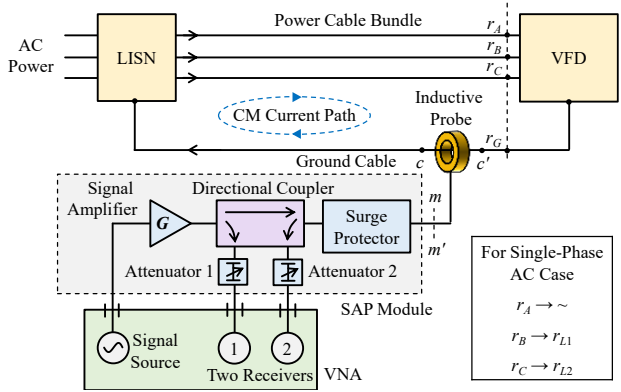

(a)

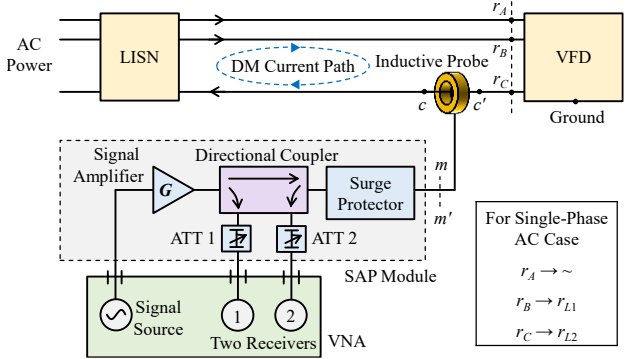

(b)

Fig. 19. FD-SPS to extract (a) $Z_{CM,VFD}$ and (b) $Z_{DM,VFD}$.

Fig. 20 shows the CM equivalent circuit of Fig. 19(a), where $V_{CM,VFD}$ represents the CM noise voltage source of the VFD system, $Z_{CM,CABLE}$ denotes the CM loop impedance formed by the power cable bundle and ground cable, and $Z_{CM,LISN}$ represents the CM impedance of the LISN. To extract in-circuit $Z_{CM,VFD}$, a stepped sweep-sine excitation is generated by the signal source of the VNA. The excitation signal is amplified by the signal amplifier, and then the amplified signal is fed to the inductive probe through a directional coupler. The incident excitation signal and the reflected signal from the inductive probe are separately sampled by the coupled ports of the directional coupler. Two attenuators are added to ensure the measured signals are within the maximum allowable levels of the VNA receivers. The receivers measure the respective directional coupler's output signals. Any high voltage transient events in the VFD system are suppressed by a surge protector to safeguard the measuring instrument. Fig. 21 shows the cascaded two-port networks representation of Fig. 20 from $m$-$m'$ [24]. $\Gamma_m$ is the reflection coefficient observed at $m$-$m'$, which is calculated directly by the VNA using the incident and reflected signals measured by the two receivers. $N_{CM,LC}$ represents the two-port network of the $Z_{CM,LISN}$ and $Z_{CM,CABLE}$. Since $N_{IP}$ and

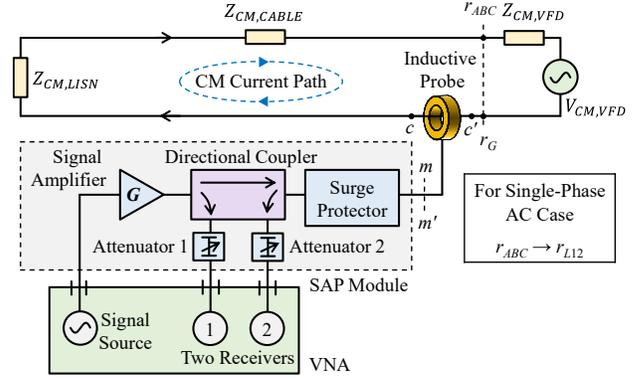

Fig. 20. CM equivalent circuit of Fig. 19(a).

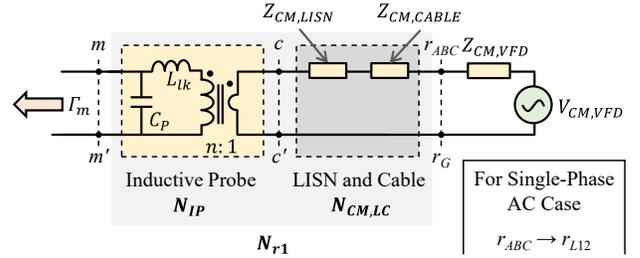

Fig. 21. Cascaded two-port networks representation of Fig. 20 from $m$-$m'$.

$N_{CM,LC}$ are cascaded, the resulting two-port network $N_{r1}$ can be expressed as:

$$N_{r1} = N_{IP} N_{CM,LC} \quad (21)$$

From Fig. 21, the relationship between $Z_{CM,VFD}$ and $\Gamma_m$ can be established in terms of the $ABCD$ parameters of $N_{r1}$:

$$Z_{CM,VFD} = \frac{k_1' \cdot \Gamma_m + k_2'}{\Gamma_m + k_3'} \quad (22)$$

where

$$k_1' = -\frac{Z_0 \cdot D_{r1} + B_{r1}}{Z_0 \cdot C_{r1} + A_{r1}}$$
$$k_2' = -\frac{Z_0 \cdot D_{r1} - B_{r1}}{Z_0 \cdot C_{r1} + A_{r1}} \quad (23)$$
$$k_3' = \frac{Z_0 \cdot C_{r1} - A_{r1}}{Z_0 \cdot C_{r1} + A_{r1}}$$

As mentioned in Section II-C, $k_1'$, $k_2'$ and $k_3'$ can be determined via executing a pre-measurement calibration. Therefore, $Z_{CM,VFD}$ can be derived from (22) based on the extracted $\Gamma_m$ using the VNA.

For in-circuit extraction of $Z_{DM,VFD}$, Fig. 22 shows the DM equivalent circuit of Fig. 19(b), in which $V_{DM,VFD}$ represents the DM noise voltage source of the VFD system, $Z_{DM,CABLE}$ denotes the DM loop impedance formed by the power cables, and $Z_{DM,LISN}$ denotes the DM impedance of the LISN. Similarly, Fig. 23 shows the cascaded two-port networks representation of Fig. 22 from $m$-$m'$, where $N_{DM,LC}$ represents the two-port network of the $Z_{DM,LISN}$ and $Z_{DM,CABLE}$. Since

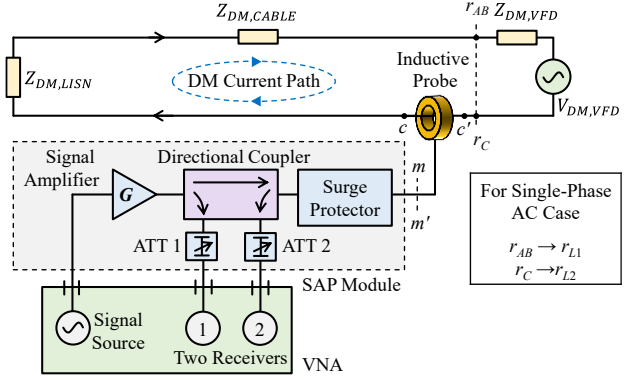

Fig. 22. DM equivalent circuit of Fig. 19(b).

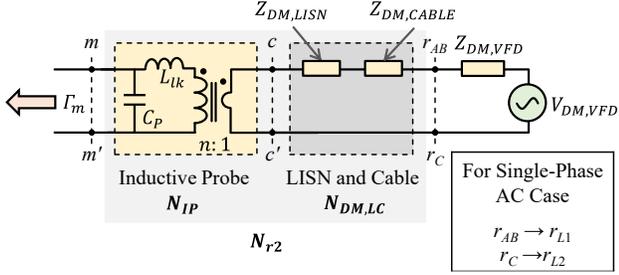

Fig. 23. Cascaded two-port networks representation of Fig. 22 from $m$-$m'$.

$N_{IP}$ and $N_{DM,LC}$ are cascaded, the resulting two-port network $N_{r2}$ can be expressed as:

$$N_{r2} = N_{IP} N_{DM,LC} \quad (24)$$

Therefore, the relationship between $Z_{DM,VFD}$ and $\Gamma_m$ can be established in terms of the *ABCD* parameters of $N_{r2}$:

$$Z_{DM,VFD} = \frac{k_1'' \cdot \Gamma_m + k_2''}{\Gamma_m + k_3''} \quad (25)$$

where

$$k_1'' = -\frac{Z_0 \cdot D_{r2} + B_{r2}}{Z_0 \cdot C_{r2} + A_{r2}}$$
$$k_2'' = -\frac{Z_0 \cdot D_{r2} - B_{r2}}{Z_0 \cdot C_{r2} + A_{r2}} \quad (26)$$
$$k_3'' = \frac{Z_0 \cdot C_{r2} - A_{r2}}{Z_0 \cdot C_{r2} + A_{r2}}$$

Similarly, $k_1''$, $k_2''$ and $k_3''$ can be determined via executing the pre-measurement calibration described in Section II-C. Therefore, $Z_{DM,VFD}$ can be derived from (25) based on the extracted $\Gamma_m$ using the VNA.

For proof of concept, a commercially available VFD system is selected as a case study. Table III lists the detailed specifications of the VFD system, LISN and cables. Table IV gives the information of each component of the FD-SPS. Based on the FD-SPS, the in-circuit $Z_{CM,VFD}$ and $Z_{DM,VFD}$ are extracted under six operating modes of the VFD to study the characteristics of $Z_{CM,VFD}$ and $Z_{DM,VFD}$ in each mode. Since the selected variable frequency controller supports voltage/frequency (V/F) and sensorless-vector (SLV) control modes, and the rated frequency of the induction motor is 50 Hz, the selected operating modes are listed in Table V.

Fig. 24(a) shows the measured $Z_{CM,VFD}$ under the six operating modes. In addition, a series of comparisons of the measured $Z_{CM,VFD}$ under each operating mode are shown in Figs. 24(b)-(f). As seen in Figs. 24(b)-(d), at the same controller output frequency, $Z_{CM,VFD}$ under different control modes (i.e., V/F control and SLV control) show good agreement from 150 kHz – 30 MHz. As seen in Figs. 24(e)-(f), at the same control mode, $Z_{CM,VFD}$ under different controller output frequencies (i.e., 10, 30, and 50 Hz) also show good agreement over most of the frequency range from 150 kHz – 30 MHz. Similarly, Fig. 25 (a) shows the measured $Z_{DM,VFD}$ under the six operating modes. A series of comparisons of the measured $Z_{DM,VFD}$ under each operating mode are also conducted and shown in Figs. 25(b)-(f). As observed in the figures, the V/F and SLV/ control modes show negligible effects on $Z_{DM,VFD}$. Likewise, at the same control mode, $Z_{DM,VFD}$ under different controller output frequencies show rather good consistency over most of the frequency range from 150 kHz – 30 MHz.

TABLE III. SPECIFICATIONS OF VFD, LISN AND CABLES

| Component | Specifications |
|---|---|
| Variable Frequency Controller | TECO L510s (No built-in EMI filter), Control mode: V/F control, SLV control Output frequency: 50 Hz |
| Induction Motor | RMS8024/B3 (3 phase, 4 pole, 0.75 kW, 50 Hz) |
| LISN | Electro-Metrics MIL 5-25/2 (100 kHz-65 MHz) |
| Cables | Controller to Induction Motor: 60 cm VFD to LISN: 100 cm |

TABLE IV. INFORMATION OF EACH COMPONENT OF FD-SPS

| Component | Details |
|---|---|
| Inductive Probe | SOLAR 9144-1N (4 kHz-100 MHz) |
| VNA | Omicron Bode 100 |
| Signal Amplifier | Mini Circuits LZY–22+ (100 kHz-200 MHz) |
| Directional Coupler | DC3010A (10 kHz-1 GHz) |
| Surge Protector | SSC-N230/01 |
| Attenuator 1 | AIM-Cambridge 27-9300-6 (6 dB) |
| Attenuator 2 | AIM-Cambridge 27-9300-3 (3 dB) |

TABLE V. OPERATING MODES OF THE VFD SYSTEM

| Operating Mode | Control Mode | Controller Output Frequency |
|---|---|---|
| Mode 1 | V/F Control | 10 Hz |
| Mode 2 | V/F Control | 30 Hz |
| Mode 3 | V/F Control | 50 Hz |
| Mode 4 | SLV Control | 10 Hz |
| Mode 5 | SLV Control | 30 Hz |
| Mode 6 | SLV Control | 50 Hz |

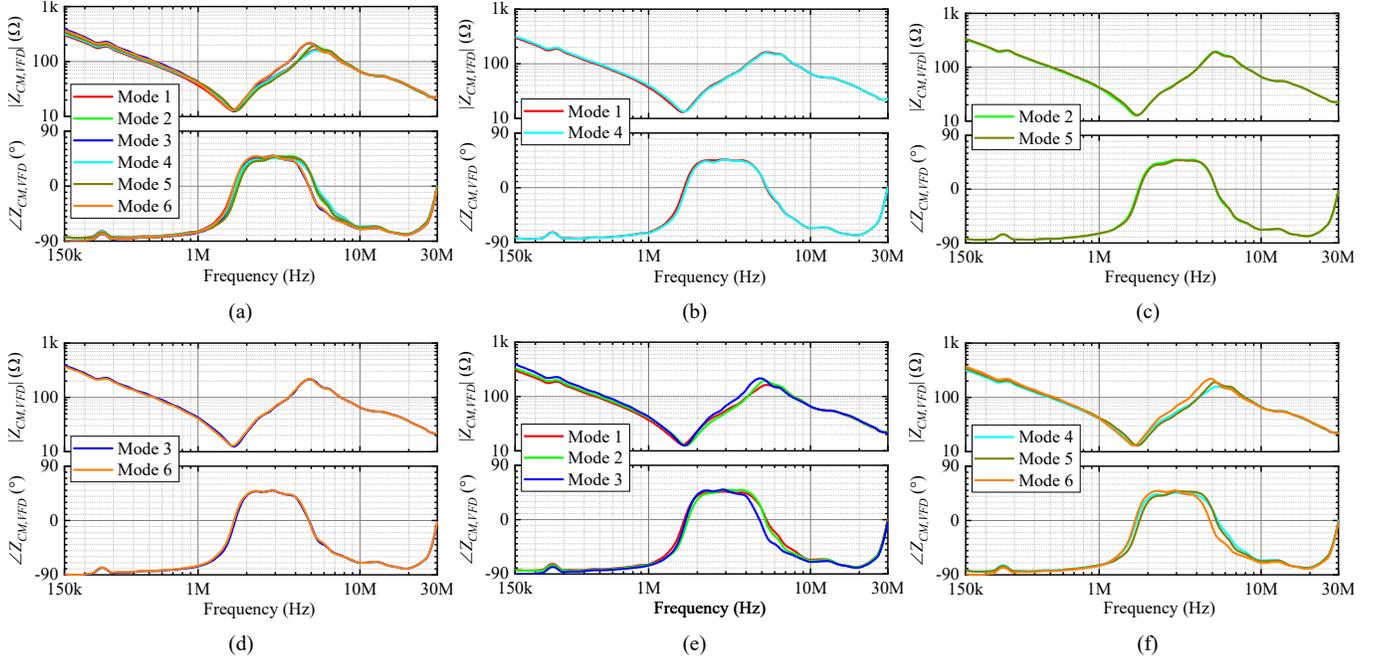

Fig. 24. Measured in-circuit $Z_{CM,VFD}$ from 150 kHz to 30 MHz under: (a) six different operating modes; (b) mode 1 (V/F–10Hz) and mode 4 (SLV–10 Hz); (c) mode 2 (V/F–30 Hz) and mode 5 (SLV–30 Hz); (d) mode 3 (V/F–50 Hz) and mode 6 (SLV–50 Hz); (e) mode 1 (V/F–10 Hz), mode 2 (V/F–30 Hz), and mode 3 (V/F–50 Hz); (f) mode 4 (SLV–10 Hz), mode 5 (SLV–30 Hz), and mode 6 (SLV–50 Hz).

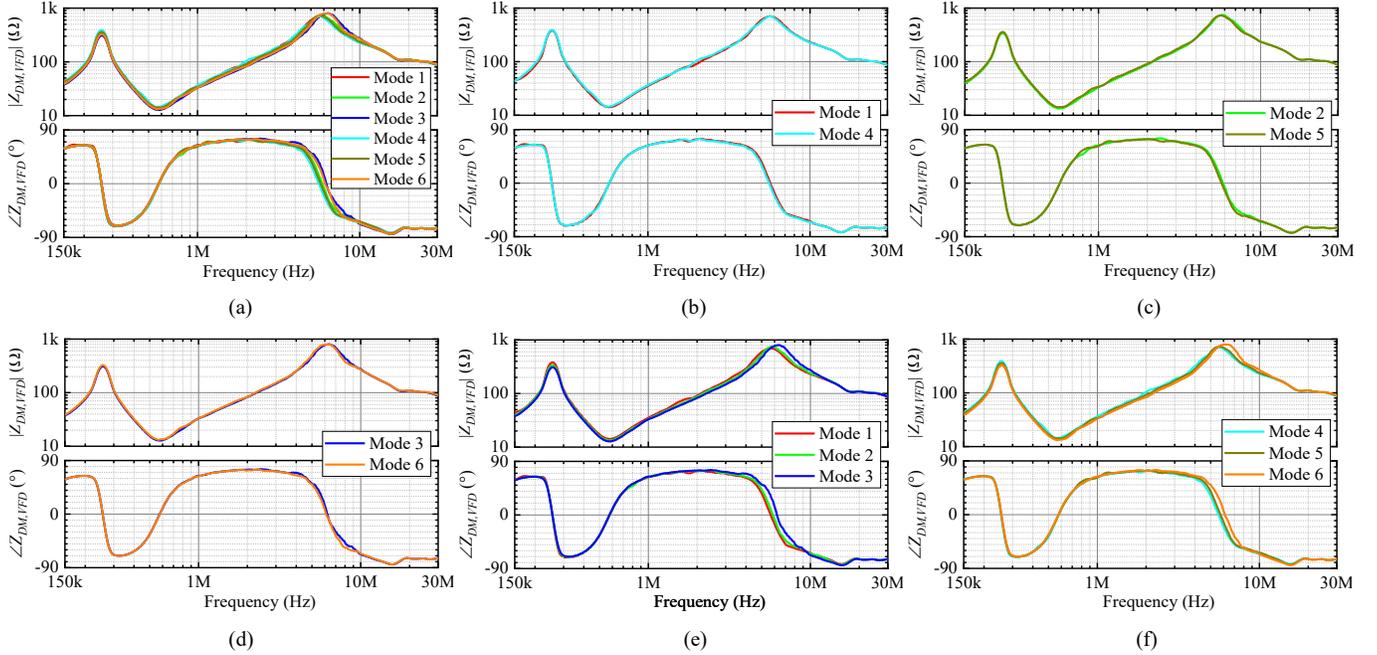

Fig. 25. Measured in-circuit $Z_{DM,VFD}$ from 150 kHz to 30 MHz under: (a) six different operating modes; (b) mode 1 (V/F−10Hz) and mode 4 (SLV–10 Hz); (c) mode 2 (V/F–30 Hz) and mode 5 (SLV–30 Hz); (d) mode 3 (V/F–50 Hz) and mode 6 (SLV–50 Hz); (e) mode 1 (V/F–10 Hz), mode 2 (V/F–30 Hz), and mode 3 (V/F–50 Hz); (f) mode 4 (SLV–10 Hz), mode 5 (SLV–30 Hz), and mode 6 (SLV–50 Hz).

## IV. CONCLUSION AND FUTURE RESEARCH

This paper presents and summaries the latest research on the inductive coupling approach for in-circuit impedance measurement as well as its EMC applications. Three common measurement setups, namely the FD-TPS, TD-TPS, and FD-SPS, are discussed, along with their respective merits and disadvantages. In addition, the various EMC applications of the inductive coupling approach are introduced, among which the radiated emission estimation of PV systems and the noise source impedance extraction at the AC input of VFD systems are elaborated. For future research, considering that all three inductively coupled measurement setups (i.e., TF-TPS, TD-

TPS, and FD-SPS) perform the measurement via stepped swept-sine or single-sine excitation, a new measurement setup with multi-sine excitation will be developed for multi-frequency simultaneous measurement of in-circuit impedance.

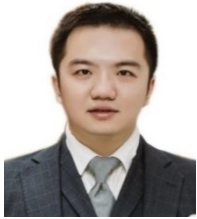

**Zhenyu Zhao** received the B.Eng. degree in electrical engineering from Huazhong University of Science and Technology, Wuhan, China, in 2015, and the M.Sc. in power engineering and Ph.D. degrees in electrical engineering from Nanyang Technological University, Singapore, in 2016 and 2021, respectively.

He is currently a Research Fellow in the School of Electrical and Electronic Engineering, Nanyang Technological University. He has authored and co-authored more than 40 refereed technical papers. His research interests include electromagnetic compatibility (EMC), impedance measurement, and health monitoring.

Dr. Zhao was a recipient of the Best Student Paper Award at the 2018 Joint IEEE International Symposium on Electromagnetic Compatibility & Asia-Pacific Symposium on Electromagnetic Compatibility (EMC & APEMC), the Best Presentation Award at the 2020 IEEE 1st China International Youth Conference on Electrical Engineering (CIYCEE), and the Best Paper Award Finalists at the 2021 Asia-Pacific International Symposium on Electromagnetic Compatibility (APEMC). He was nominated and invited to participate in the Global Young Scientists Summit (GYSS) in 2019 and 2022. He has served as the Session Chair, a TPC Member and an Invited Speaker for several international conferences and seminars. Since 2022, he has been serving as the Secretary for the IEEE EMC Singapore Chapter.

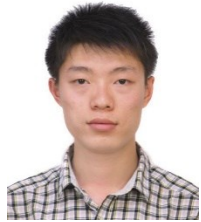

**Fei Fan** received the B.Eng. degree in electrical engineering from Tianjin University, Tianjin, China, in 2014, and the M.Sc. in power engineering and Ph.D. degrees in electrical engineering from Nanyang Technological University (NTU), Singapore, in 2015 and 2020, respectively.

He is currently a Research Fellow in the School of Electrical and Electronic Engineering, NTU. His research interests are electromagnetic compatibility and electromagnetic interference measurement, in-circuit impedance extraction, and EMI filter design for motor-drive system.

Dr. Fan is the recipient of the Best Student Paper Award at the Asia-Pacific Symposium on Electromagnetic Compatibility (APEMC) 2017 and Progress in Electromagnetics Research Symposium (PIERS) in 2017.

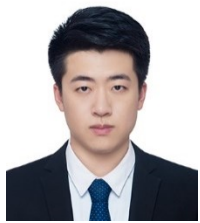

**Huamin Jie** received the B.Eng. degree in electrical engineering from Wuhan University, Wuhan, China, in 2019, and the M.Sc. degree in power engineering from Nanyang Technological University, Singapore, in 2020, respectively. He is currently working toward the Ph.D. degree with the School of Electrical and Electronic Engineering, Nanyang Technological University.

His research interests include impedance measurement, device modeling, electromagnetic interference (EMI), and EMI filter design.

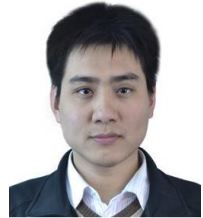

**Quqin Sun** received the B.Eng. and Ph.D. degrees in electrical engineering from Huazhong University of Science and Technology, Wuhan, China. in 2011 and 2016, respectively.

He worked as a Research Associate at the Institute of Fluid Physics of China Academy of Engineering Physics till 2018. From 2018 to 2021, he joined Nanyang Technological University, Singapore as a Research Fellow. He is now an Engineer in Wuhan Second Ship Design and Research Institute. His research interest spans within high-field pulsed magnet design, electromagnetic launch, laser-ultrasonic non-destructive inspection and motor condition monitoring.

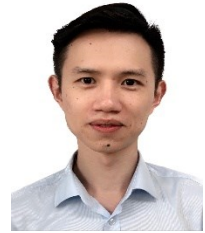

**Pengfei Tu** received the B.Eng. degree in electrical engineering from Wuhan University, Wuhan, China, in 2013, and the M.Sc. and Ph.D. degrees in power engineering from Nanyang Technological University, Singapore, in 2014 and 2019, respectively.

He is currently a Research Fellow in the School of Electrical and Electronic Engineering, Nanyang Technological University. His current research interests include reliability of multilevel converters, model predictive control, wireless power transfer and energy management system.

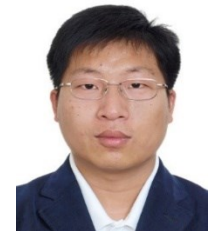

**Wensong Wang** received the Ph.D. degree in Communication and Information Systems from Nanjing University of Aeronautics and Astronautics, Nanjing, China, in 2016.

From 2013 to 2015, he was a Visiting Scholar with the University of South Carolina, Columbia, USA. In 2017, he joined Nanyang Technological University, Singapore, as a Research Fellow, and now he is a Senior Research Fellow.

His research interests include MIMO antenna, low-frequency (coil) antenna, and advanced sensors.

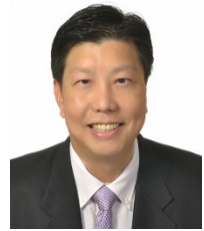

**Kye Yak See** received the B.Eng. degree in electrical engineering from National University of Singapore, Singapore, in 1986, and the Ph.D. degree in electrical engineering from Imperial College London, U.K., in 1997.

From 1986 and 1991, he was with Singapore Technologies Electronics, Singapore, as a Senior Engineer. From 1991 to 1994, he was a lead Design Engineer with ASTEC Custom Power, Singapore. Since 1997, he has been with Nanyang Technological University (NTU), Singapore, as a Faculty Member. He is currently an Associate Professor with the School of Electrical and Electronic Engineering, NTU. He holds concurrent appointment as Director of the Electromagnetic Effects Research Laboratory and Director of SMRT-NTU Smart Urban Rail Corporate Laboratory. His current research interests are electromagnetic compatibility (EMC), signal integrity and real-time condition monitoring.

Dr. See was the Founding Chairs of the IEEE Electromagnetic Compatibility (EMC) Chapter, IEEE Aerospace and Electronic Systems, and the IEEE Geoscience and Remote Sensing Joint Chapter in Singapore. He was the General Chairs of 2015 Asia Pacific Conference on Synthetic Aperture Radar (APSAR 2015) and 2018 International Conference on Intelligent Rail Transportation (ICIRT 2018). Since January 2012, he has been the Technical Editor of the IEEE Electromagnetic Compatibility Magazine.